\documentclass{article}

\usepackage[english]{babel}
\usepackage{amsmath}
\usepackage{fullpage}
\usepackage{graphicx}
\usepackage{subfigure}
\usepackage{xspace}
\usepackage{color}
\usepackage{caption}
\usepackage{natbib}

\def \rank{\mathop{\rm rank}\limits}

\title{Supervised Penalized Canonical Correlation Analysis}
\author{Andrea Thum, Susann M\"onchgesang, Lore Westphal, Tilo L\"ubken, Sabine Rosahl,\\ Steffen Neumann and Stefan Posch}

\begin{document}
\maketitle

\begin{abstract}
{\bf Motivation:}
The canonical correlation analysis (CCA) is commonly used to analyze data sets with paired data, e.g. measurements of gene expression and metabolomic intensities of the same experiments.
This allows to find interesting relationships between the data sets, e.g. they can be assigned to biological processes. However, it can be difficult to interpret the processes and often the relationships observed are not related to the experimental design but to some unknown parameters. \\

{\bf Results:}
Here we present an extension of the penalized CCA, the {\em supervised penalized} approach (spCCA), where the experimental design is used as a third data set and the correlation of the biological data sets with the design data set is maximized to find interpretable and meaningful canonical variables. 

The spCCA was successfully tested on a data set of {\em Arabidopsis thaliana} with gene expression and metabolite intensity measurements and resulted in eight significant canonical variables and their interpretation. We provide an R-package under the GPL license.\\

{\bf Availability:}
R package spCCA at http://msbi.ipb-halle.de/msbi/spCCA/\\

{\bf Contact:} 
andrea.thum@informatik.uni-halle.de

\end{abstract}

\section{Introduction}

Systems biology aims to understand living organisms, often by
combining multi-factorial experiments and multiple assay techniques to
obtain, e.g., gene expression, protein or metabolite levels. 
To unravel the interactions between genes, proteins, or metabolites,
statistical methods are used to discover dependencies among the data.

Many genes are already known to be involved in the control of metabolism and activation 
of pathways. Correlations between genes
and specific metabolites have been used to assign signaling functions
to the metabolites (\citet{Hannah10}). Moreover, genes encoding
enzymes for secondary metabolite synthesis have been identified by
specifically looking for expression profiles of possible candidate
genes (\citet{Muroi09}).

In order to detect linear correlations between two data sets, the canonical correlation analysis (CCA, \mbox{\citet{Hotel36}}) can be used. 
The CCA returns a pair of linear combinations of the features (e.g. gene or metabolite levels) in each of the two data sets, which correlate maximally: the first pair of canonical variates, which is the first canonical variable. Orthogonal to this pair, the second pair of combinations with second largest correlation coefficient can be found, and so forth.

The canonical variables for the observed biological data contain information about the underlying processes in the organism, where
a large weight for one feature in the linear combination corresponds to large influence of this feature to the process.
A process can be deduced from the pattern of the course of the canonical variable across the samples.

For large-scale data sets, there is often an imbalance between the large number of features and the  
much smaller number of biological samples measured. 
A robust and sparse solution is neccessary to deal with the underdetermination and to extract
the most relevant features. This can be achieved by
penalized CCA (pCCA, \citet{Waaij09}), where an elastic net solution is implemented which combines two penalty terms. The ridge regression term is used to eliminate the singularity. A lasso regression term is implemented which forces small weights within the linear combinations to zero, thus essentially removing them.

A biological system is influenced by many factors, which can be caused by the experimental design, but also by parameters beyond the control of the experimentalist. Often it is difficult to recognize which processes are associated with the canonical variable as there may be processes that interfere with each other resulting in a complicated pattern of the canonical variable.
Furthermore, processes independent from the experimental design are difficult to interpret. 
Therefore, it is desirable to associate the canonical variables with the experimental design.

We extend the pCCA to more than two data sets and use the experimental design as an additional data set to obtain a {\em supervised  penalized} CCA (spCCA).
In this case, the sum of the correlation coefficients between the canonical variables of each biological data set with the experimental design data set is maximized. Thus, a high correlation can be achieved between the linear combination of the features of each biological data set and the linear combination of the vectors of the experimental design. These combinations can be interpreted easily in terms of the experimental design.

The paper is structured as follows: in Section 2.1 we first introduce standard and penalized canonical correlation analysis for two data sets, as well as the generalized CCA for more than two data sets. 
In 2.2 we show how we combined generalized and penalized CCA and adopted this approach to obtain the new supervised penalized CCA. Section 2.3 gives details of the biological experiments including two assays 
 obtained from {\em Arabidopsis thaliana}, where both gene expression 
and metabolite levels were determined.
In the results section 3 we apply the new spCCA
to these data sets. We identify several well-explainable processes,
and compare the results to standard pCCA to demonstrate the potential of the supervised approach.

\section{Materials and methods}

We consider $n$ experiments with two sets of features.
The features in the sets are generally of different type, e.g., metabolite intensities and gene expression. 
The  $p_1$ and $p_2$ features are collected in a $n \times p_1$-matrix $X_1$ and a $n \times p_2$-matrix $X_2$,
where rows represent the experiments. The variables of the matrices are normalized columnwise to have zero means and unit variance.

\subsection{Canonical Correlation Analysis}

In order to maximize the correlation between the linear combinations $X_1{w_1}$ and $X_2{w_2}$ of the columns of the matrices $X_1$ and $X_2$ the CCA determines weight vectors ${w_1}$ and ${w_2}$.
The resulting linear combinations are the pair of canonical variates for the first canonical variable.

\newpage

\subsubsection{Standard CCA}

To compute the standard CCA, the correlation coefficient of the linear combinations is to be maximized with respect to $w_1$ and $w_2$ :
 
\def\corr{\operatorname{corr}}

\begin{align}
 \corr(X_1w_1, X_2w_2) = \frac{w^T_1X_1^TX_2w_2}{\sqrt{w_1^TX_1^TX_1w_1w_2^TX_2^TX_2w_2}}\rightarrow \max,\notag
\end{align}

where  $X_1^TX_1$  and $X_2^TX_2$ are the variance matrices and $X_1^TX_2$ is the co-variance matrix of data sets $X_1$ and $X_2$.

This is equivalent to:
\begin{align}
w_1^T{X_1}^TX_2w_2 \rightarrow \max\notag
\end{align}
with constraint:
\begin{align}
 w_1^T{X_1}^TX_1w_1 =  w_2^T{X_2}^TX_2w_2 = 1 \notag
\end{align}

This leads to a generalized eigenvalue problem, where the maximal eigenvalue corresponds to the correlation coefficient between the first canonical variates for $X_1$ and $X_2$, and the corresponding eigenvector yields the weights of the canonical variate for data set $X_1$. The weights for $X_2$ can be inferred (\citet{Hardoon03}). 
The weights of the canonical variates indicate the contribution of each
feature, e.g. the gene or metabolite, to the correlation.

So far we considered the combination with the highest  correlation coefficient achievable. Further combinations of features with lower correlation can be inferred from the remaining eigenvectors. 
There are $\min(\rank(X_1),\rank(X_2))$ canonical variables, which are orthogonal to each other with decreasing correlation coefficients.

\subsubsection{Penalized CCA}

In data sets from biological experiments the number of experiments is often much smaller than the number of features. For data matrices with $p_1>n$ or $p_2>n$ , the variance matrix is singular and thus not invertible, and the CCA-problem is ill-posed. 
\citet{Park09} as well as \citet{Waaij09} propose the pCCA solution to address this problem by incorporating the elastic net approach. 

The elastic net (\citet{Zou05}) is a combination of two regression penalties: ridge regression and lasso. Ridge regression is implemented by regularizing a matrix $M$, to make them full rank and thus invertible (\citet{Hastie09}). For this purpose, $\gamma I$ is added to the matrix, where $\gamma$ is a positive scalar parameter and $I$ is the identity matrix. 

Since not all features are expected to be involved in the underlying process, the lasso penalty aims at eliminating unimportant features. Weights less than a parameter $\frac 1 2 \lambda$ are set to zero (\citet{Tib96}).
To avoid a double shrinkage by this two-stage procedure the coefficients are multiplied by $(1+\gamma)$. 

The elastic net can deal with data sets with more features than experiments, produces sparse results and shows a \lq grouping effect\rq, i.e. it assigns similar weights to highly correlated features within each data set. 

The elastic net has no analytical solution. To integrate this approach into the standard CCA framework, \citet{Waaij09} translated this problem into a coupled regression framework. This framework can be solved iteratively by the power method, which determines the eigenvectors $w_1$ and $w_2$ for the canonical variates $X_1w_1$ and $X_2w_2$ for the dominant eigenvalue. 
To compute the next pair, which is orthogonal to the first, the data sets $X_k^{'}$ are constructed orthogonally to the first variable  by the subtraction of the canonical variates from the data sets. This can be done by regressing each column of $X_k$ on the canonical variate $X_kw_k$ and keeping the orthogonal residual (\citet{Waaij09}).

For each data set, two parameters are required: the ridge regression parameter $\gamma_k$ to control the strength of regularization and lasso parameter $\lambda_k$ for the sparsity. It is not obvious how to set these parameters. 
\citet{Park09} suggest a strong ridge regression regularization and set the ridge regression parameter to infinity.  The lasso parameters  are determined via resampling, maximizing the test sample correlation.

\subsubsection{Generalized CCA}

As we aim to analyze more than two data sets by a penalized CCA we consider first the standard generalized canonical anlysis.

The standard generalized canonical correlation analysis (gCCA) computes the CCA for $m>2$ data sets with $n$ experiments and $p_k$ features, $k=1,\dots,m$. Here, different optimization criterions are possible. We used 
the sum of the correlation coefficients of the canonical variates between all pairs $X_k$, $X_l$ of the data sets, which is to be maximized (SUMCOR formulation in \citet{Kett71}):
\begin{align}
\frac {1}{m(m-1)} \sum_{k,l=1,\atop k\ne l}^m w_k^TX_k^TX_lw_l \rightarrow \max\notag
\end{align}
with constraint:
\begin{align}
w_k^TX_k^TX_kw_k = 1, ~ k=1, \dots m \notag
\end{align}

For the standard CCA, a canonical variable is a pair of two maximally correlated variates. Now, a generalized canonical variable for $m$ data sets consists of $m$ variates, where the sum of the correlation coefficients of all pairs of these variates is maximal.

The SUMCOR optimization problem has no closed form solution.
The eigenvalue problem can be translated into a  regression framework.
As described in \citet{Via06} this results in $m$ coupled regression problems, which can be solved iteratively.

For iteration step $t$ this yields:
\begin{align}
w_{k}^{(t)} =  X_k^+\cdot\sum_{i=1}^mX_iw_{i}^{(t-1)}, ~k=1, \dots , m\notag
\end{align}
~ where $X_k^+=(X_k^TX_k)^{-1}X_k^T$ is the pseudoinverse of $X_k$. The $w_{k}^{(t)}$ have to be normalized to length $1$ before the next iteration step.

\subsection{Supervised pCCA}

In addition to $m-1$ biological data sets $X_1, \dots, X_{m-1}$, we include the design data set $X_m$ to the CCA. This design data set is semantically similar to the design matrix of the experiments and describes the experimental setup, for example the growth condition, mutations or treatment.
The information about the experimental design can be encoded in binary design vectors of size $n$, the number of experiments. These vectors describe the group membership of each experiment to different experimental conditions.  
Depending on the experiments this yields $p_m$ vectors each in analogy to the measurements of one feature in all $n$ experiments. 
For example, the feature vector for $n=10$ and a treatment vs. control setup with  five replicates each is given as: $(0~0~0~0~0~ 1~ 1~ 1~ 1~ 1)^T$.
The combination of these vectors as columns yields the $n \times p_m$ design matrix $X_m$.

We call the generalized pCCA applied to these $m$~data sets \textit{supervised penalized CCA (spCCA)}
as knowledge about the experimental setup is directly intergrated and exploited.
The weights  of the canonical variate $X_m w_m$ yield a combination of the experimental conditions and facilitate the interpretation of the underlying processes.

If the sum of the pairwise correlation coefficients is maximized as proposed by the SUMCOR approach, high correlation coefficients between pairs of biological data sets might compensate low correlation between biological data sets and the design data set. Thus, as we are especially interested in a high correlation with the design data set, we only consider the correlation coefficients for each biological data set with the design data set. This does not take the correlation between the biological data sets into account and maximizes the following problem:
\begin{align}
\frac {1}{m-1} \sum_{k=1}^{m-1} w_k^TX_k^TX_mw_m \rightarrow \max\notag,\\
w_k^TX_k^TX_kw_k = 1, ~ k=1, \dots m \notag
\end{align}
 where $X_m$ denotes the design data set.

This results in a reduced iterative solution:
\begin{align}
w_{k}^{(t)} &= w_k^{(t-1)} +X_k^+ X_mw_{m}^{(t-1)} , ~k=1, \dots, m-1\notag\\
w_{m}^{(t)} &= w_m^{(t-1)} +X_m^+\cdot\sum_{i=1}^{m-1}X_iw_{i}^{(t)}\notag
\end{align}
We used a strong regularization for the elastic net, which sets the co-variances in the pseudoinverses $X_k^+$
of each data matrix to zero (\citet{Park09}). 

The lasso penalty is included by setting weights below a threshold $\frac 1 2 \lambda_k$ to zero in each iteration step.

It is difficult to adjust the lasso parameters $\lambda_k$  to adequately control the sparsity of the penalized CCA. 

We used a resampling technique to determine the parameters. Using a grid search on the $\lambda_k$ the following algorithm for the spCCA is repeated several times for different training data sets and for each combination of $\lambda_k$. For our Arabidopsis data set, ten training data sets were sampled. For each training data set, one eighth of the experiments was randomly drawn.

\newpage
{\bf ALGORITHM FOR spCCA}
\begin{enumerate}
\item INPUT: $m-1$ biological data sets $X_1, \dots, X_{m-1}$, normalized to zero mean and unit variance; one design data set $X_m$, normalized to zero-mean and unit variance, extracted from the experimental design; $m$ sparsity parameters $\lambda_k$
\item Compute strong regularized pseudoinverses for the biological data sets $X_1^+, \dots , X_{m-1}^+$
\item Compute pseudoinverse of the design data set $X_m$. $X_m$ does not have to be regularized, if there are no redundant vectors.
\item Set initial normalized values for $w_{k}^{(0)}, ~k=1, \dots, m$
\item In iteration step $t$:

	\begin{enumerate}
	\item  for $1\le k<m$ (biological data sets):\\
    \begin{align}v_{k} = w_k^{(t-1)} + X_k^+ X_mw_m^{(t-1)}\notag\end{align}
     for $k=m$ (design data set):\\
     \begin{align} v_{m} =&  w_m^{(t-1)} + X_m^+ \cdot\sum_{i=1}^{m-1}X_iw_i^{(t)}
    \notag\end{align}
	\item Normalize $v_{k} = \frac {v_k} {|v_k|}$, $k=1,\dots,m$
	\item Set all components of each $v_{k}$ to zero, which are smaller or equal to the sparsity parameter $\frac {1}{2} \lambda_{k}$ ($\mathrel{\widehat{=}}$ lasso penalty)
	\item Normalize again to obtain updated weigth vector $w_k^{(t)} = \frac {v_k} {|v_k|}$, $k=1,\dots,m$
    \item if convergence criterium reached: break
    \end{enumerate}

\end{enumerate}

Due to the lasso penalty in step (c), which enforces the sparseness, the algorithm converges to an eigenvector. Depending on the initial weights, it does not necessarily converge to the vector associated with the dominant eigenvalue. 
Thus we repeat the iteration in step~5. for different initializations with random values (step~4.) ten times and keep the solution
with the largest eigenvalue for the training data set. The eigenvectors with the median eigenvalue of all training data sets is 
used as the weight vectors for the canonical variable.

Again, this determines  only  the first canonical variable. To compute further variables the variates are subtracted from the biological data sets $X_k$ to produce orthogonal data sets. 
The design data set remains unchanged.

\subsubsection{Significance of correlation}

To decide whether a canonical variable for three data sets is significant,
a permutation test was used. 
We found that for our data set the correlation coefficient needs to exceed $0.605$ 
for a level of significance $\alpha=0.05$ and needs to be larger than $0.635$  for $\alpha=0.01$.

\subsection{Data}

We demonstrate the power of the supervised pCCA approach using data from an
experiment on the response of the model plant {\em Arabidopsis
  thaliana} to the pathogen {\em Phytophthora infestans}. 
The data set consists of microarray gene expression data and 
LC/MS based metabolite profiles.

The oomycete {\em P. infestans} is the causal agent of late blight, the most
devastating potato disease. In contrast to potato, {\em A.
thaliana} is able to successfully prevent colonization of the pathogen
due to a multi-layered nonhost resistance. 

Several mutants
have been isolated which are impaired in penetration resistance.
A mutation in the gene {\em PEN2}, which encodes an enzyme involved in
indole glucosinolate metabolism (\citet{Bed09}), results in the loss
of penetration resistance against {\em P. infestans}
(\citet{Lipka05}). Despite its ability to penetrate epidermal cells of
{\em pen2} mutant plants, {\em P. infestans} is still not able to colonize these
plants. Additional mutants were isolated by \citet{Kopischke13Impairedsterolester} which show enhanced
defense responses upon infection with {\em P. infestans}: {\em pen2erp1} and {\em pen2erp2}, and 
backcrossed mutants {\em erp140} and {\em erp2D}.
 
We used six different plant lines, the wildtype-like {\em gl1},
and the five different mutants ({\em pen2, pen2erp1, pen2erp2, erp2D, erp140}).
The plants were either infected with
{\em P. infestans} spores or treated with water as control, and harvested
6h and 12h after treatment. The experiment was repeated three times with different {\em P. infestans} cultures,
resulting in biological triplicates, for an overall of $6 \times 2
\times 2 \times 3 = 72$ samples.

These samples from the same plant material were used on $72$ Affymetrix microarrays and for LC/MS-metabolite profiling (see supplemental material for details). 
For each of the $72$ samples, three LC/MS-runs were performed.
We obtained 202 LC/MS measurements (72 samples with one to three technical
replicates each) of the abundance of polar metabolites. We
used the centWave algorithm~(\citet{tautenhahn}) to extract features from
the LC/MS raw data, and used xcms~(\citet{Smith06}) to group them into a
rectangular data matrix. The technical replicates were averaged and the
metabolomic data resulted in a $ 72\times 5896$ data matrix. The metabolomic data was reduced to $3007$ putative pseudomolecular ions with help of the R-package CAMERA (\citet{CAMERA}). 
The microarray data was processed and normalized with the R-package
simpleaffy (\citet{Wilson05}) and resulted in a $72\times 22810$ data matrix. 

We reduced the data sets by excluding features with low variance (threshold chosen $\sigma<1$
for genes and $\sigma<0.4$ for metabolites), resulting in a 
$72 \times 1277$ gene expression matrix, and a $72 \times 252$ LC/MS signal intensity matrix.

\section{Results}

\begin{figure}[htbp]
    {\includegraphics[width=0.7\textwidth]{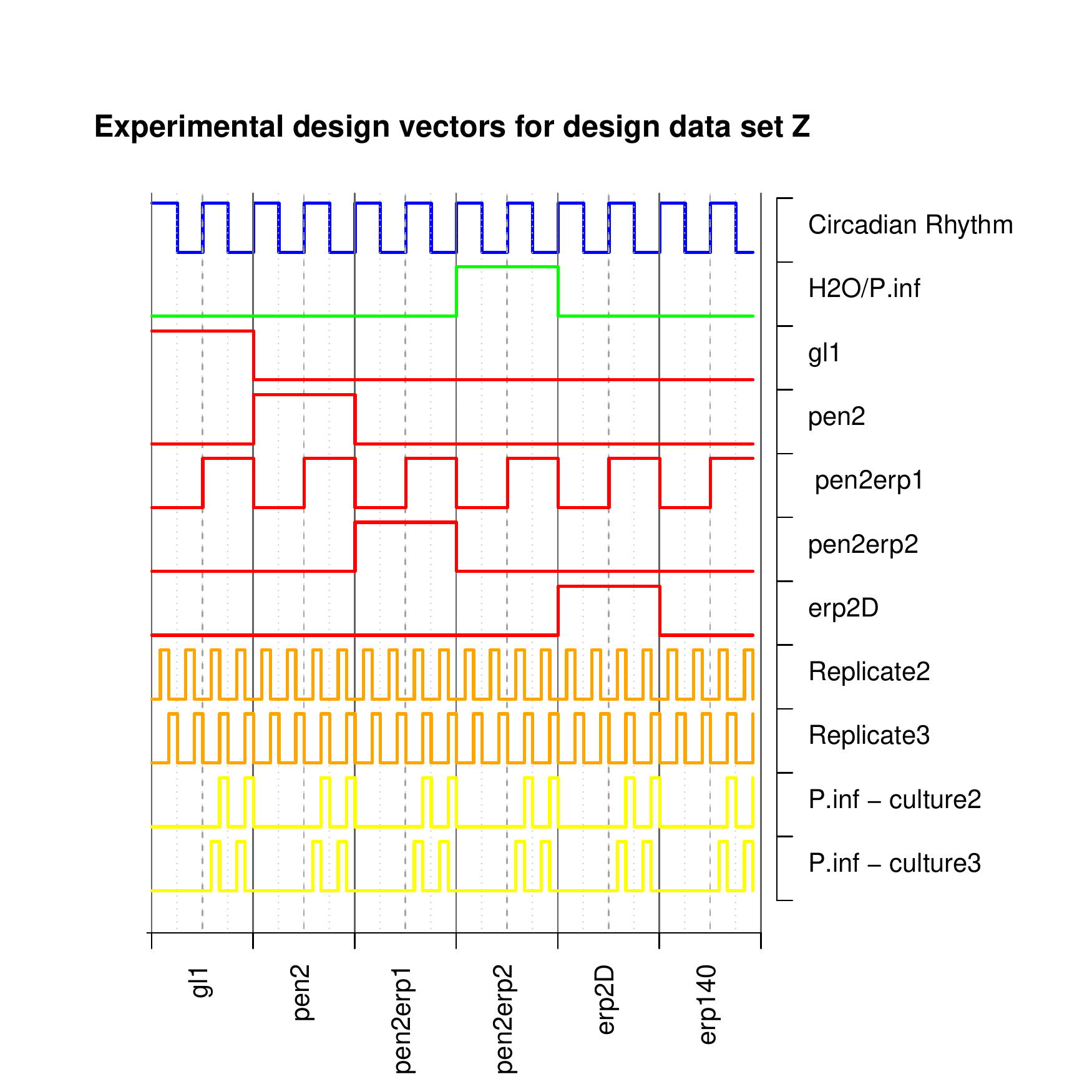}}
     \caption{\footnotesize The experimental design vectors for the 72 experiments. The biological replicates are consecutive, the solid vertical  lines separate the six mutants, the dotted lines seperate the time points (6h and 12h) and the dashed lines the treatment (H$_2$O and {\em P. infestans}).  }
	\label{FigExpDes}
 \end{figure}

The reduced data sets were analysed by the supervised pCCA.
The supervised solution included the experimental information in Fig. \ref{FigExpDes} as the design data set.

To describe six mutants non-redundantly only five vectors are required, likewise only two vectors to code three replicates and the oomycete cultures. 
In consequence, the design data set contains 11 design vectors of length 72 for the 72 experiments.

To determine the sparsity parameters for our data set we performed a ten-fold repeated hold-out sampling with a grid of size $16 \times 21 \times 31 = 10416$. This requires 
between 20 to 120 minutes using an AMD Athlon 64 Processor 4000+ with 2400 MHz and 2 GB RAM.

\begin{figure}[ht!]
    {\includegraphics[width=0.7\textwidth]{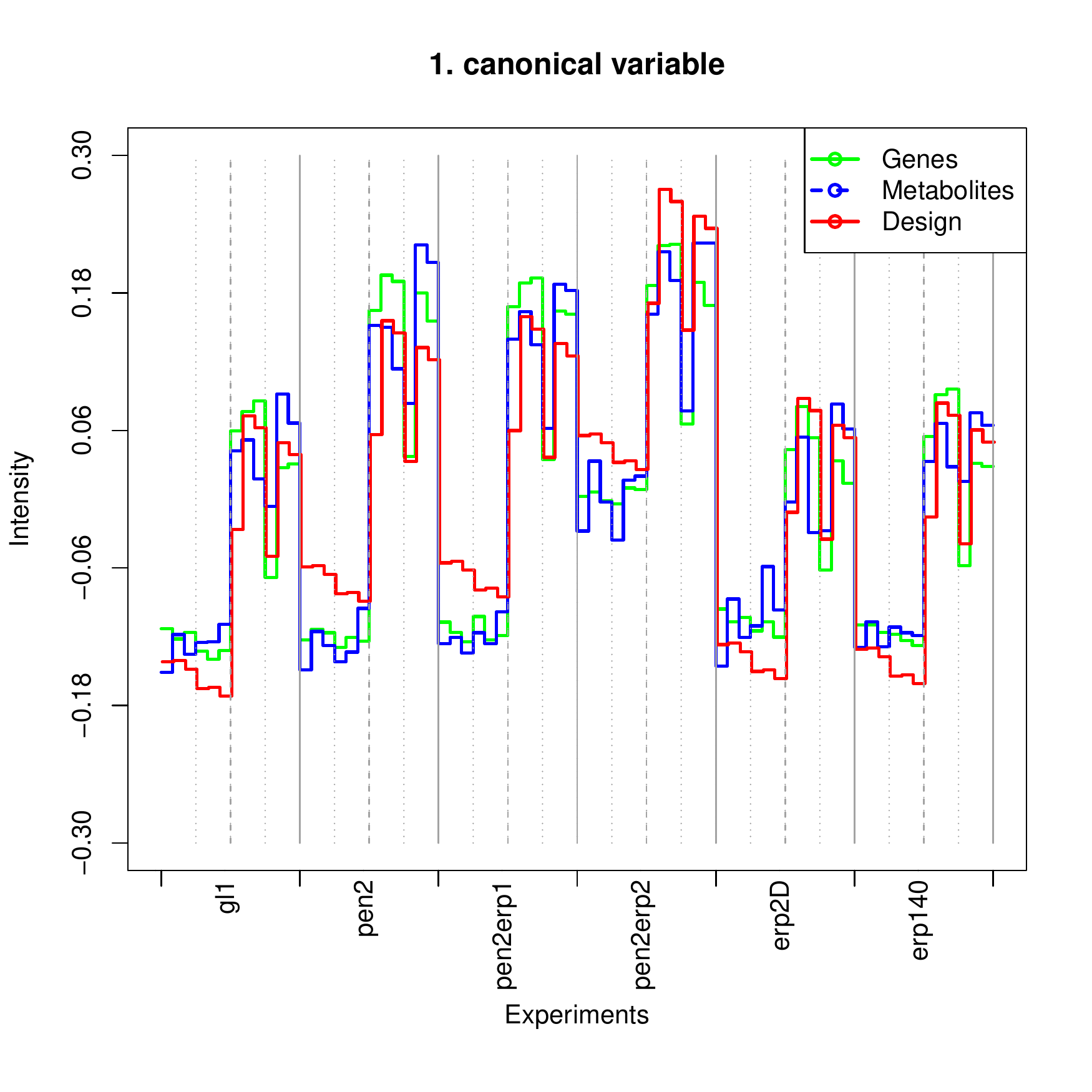}}
     \caption{\footnotesize  Supervised pCCA results: First canonical variable for the genes ($X_1w_1$), metabolites ($X_2w_2$) and the experimental design data set ($X_mw_m$) for all 72 experiments. 
}
\label{Fig3D1}
 \end{figure}

The first canonical variable (Fig. \ref{Fig3D1}) is a combination of  genotype and pathogen response, which is elevated in all three {\em pen2}-mutants
upon infection. One {\em pen2}-mutant shows exceedingly high expression values and metabolite intensities for both treatment and control.
The largest weighted metabolites include 
camalexin as well as flavonoids, which play a role in defense. This is in agreement with biological knowledge: the plant senses the attack by  {\em P. infestans} and immediately releases camalexin as first defense reaction.

The genotype {\em pen2erp2} is the explanation of the second canonical variable (for this and all further variables: see supplemental material). The metabolites salicylic acid glucoside and dihydroxybenzoic acid receive the largest weights for this variable. 
Salicylic acid appears to be present at constitutively high levels in {\em pen2erp2}.

The third variable  resembles the circadian rhythm, 
since the sample collection time was 6h and 12h after inoculation (at noon and late afternoon, respectively). 
Not surprisingly, a large number of genes show the circadian rhythm, but only a few metabolites in our data set. This might be due to the fact that the  metabolites associated with circadian rhythm (especially primary metabolism and sugars), cannot be detected by LC/MS. 
A combination of two mutations 
is found in the fourth variable: the {\em erp2}-mutation and the {\em pen2}-mutation. As the {\em pen2erp2}-component was already subtracted in the second variable, mainly mutant {\em erp2D} is increased and {\em pen2} and {\em pen2erp1} are decreased in the canonical variable. 
The gene {\em PEN2} 
can be found among the largest weighted features.

 An interesting effect is quality control of the experiment. Despite the best efforts to sustain reproducible experimental
conditions, the fifth, the sixth and the ninth canonical variable could possibly be explained by unknown environmental
factors or slight variations in the sample processing procedures 
which influenced the replicates, as well as the influence of different oomycete cultures to the plants. The spCCA allows to decompose these effects.

There are two further canonical variables (seventh and eighth) with a quite low but still significant correlation coefficient. 
They suggest a similarity between the mutants {\em pen2erp1}, {\em pen2} and {\em erp2D}, as well as differences between the mutants {\em pen2} and {\em pen2erp1}.

 \begin{figure}[ht!]
    {\includegraphics[width=0.7\textwidth]{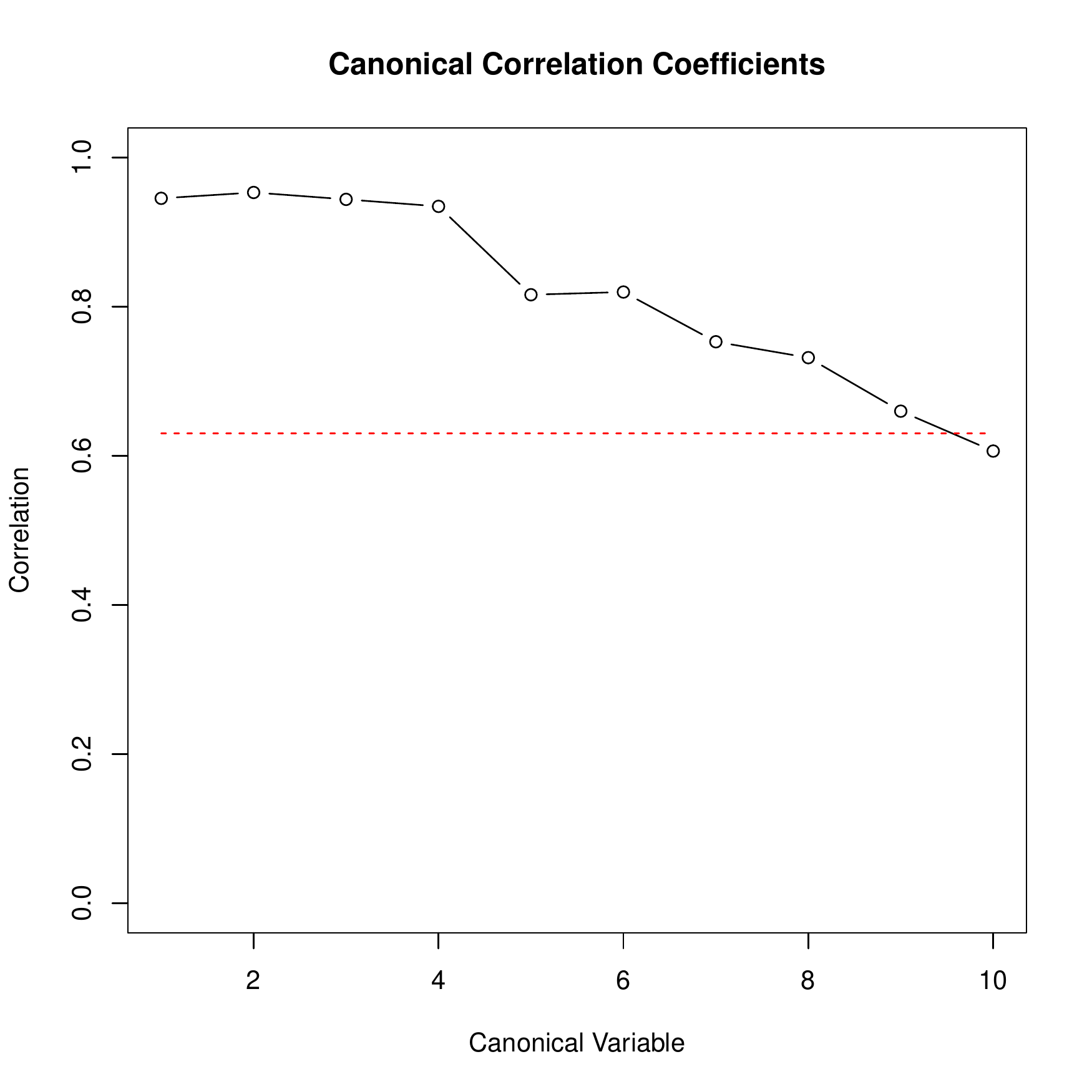}}\hfill
    \caption{\footnotesize The correlation
      coefficients for the first ten canonical variables. The dashed line indicates a significance of $\alpha=0.01$. }
	\label{FigCorrLine}
 \end{figure}

Our supervised approach was able to recognize nine significant variables for significance level $\alpha=0.01$ and to give an explanation to them. 
The main explainable processes in the plants seem to be the defense against the oomycete, as well as the circadian rhythm. The effects of two of the mutations, as well as influences of the replicates show a slightly lower correlation.
Further variables are not significantly correlated (Fig. \ref{FigCorrLine}).

For the standard CCA, which does not use sparse weight vectors, the correlation coefficients of the canonical variables are monotonously decreasing. This may not be the case in penalized CCA.
This effect can be seen in Fig. \ref{FigCorrLine}, where the correlation coefficients for the first ten variables are shown. The second and the sixth correlation coefficient are larger than their respective predecessors.\\

If standard non-supervised penalized CCA is used, ten significant combinations of genes and metabolites were found, but only two were easy to interpret.

The first canonical variable (see supplemental material) is similar to the first variable of the spCCA and shows the reaction to the infection, which is increased for the {\em pen2}-mutants.
The second variable  corresponds to mutant {\em pen2erp2}, and the main metabolite is salicylic acid.
A number of further canonical variables are found which are difficult to interpret. 
The canonical variables for the {\em pen2}-mutant as well as the {\em erp1}-mutation (fourth supervised pCCA variable) were not found.

\section{Discussion}

Discovery and interpretation of complex relationships between gene activity and metabolites is still a challenge in systems biology. A penalized canonical correlation analysis is a useful tool for this purpose. Still, the main question to a canonical variable is: which process does it resemble, what does it biologically  mean? Although some processes are easy to identify, most of the canonical variables are difficult to interpret. One solution is to check the corresponding genes and metabolites of the canonical variable -- but this is elaborate and very complicated since many genes and gene functions are still unknown, or the metabolites might not yet be identified.

Furthermore, many correlations are based on an unknown genotype effect or unobserved growth conditions. These effects are usually not interesting for the experimentalist and it is hard or even impossible to interpret the pattern. A standard pCCA can only search for high correlation, and thus well-explainable variables with lower correlation coefficient will be missed. 

The supervised pCCA provides additional information as it explicitly incorporates the design of the experiment into the  analysis, and thus the  underlying biological questions. This assists in interpreting the biological processes and guides the spCCA to find interpretable processes of interest.
We showed that this method was very useful to unravel relevant relationships between two data sets of gene expression and metabolite levels of {\em Arabidopsis thaliana} subjected to pathogen infection. 
To extract further processes  a standard penalized CCA can be applied subsequently 
  to unveil additional correlation in the residual data sets.

In the analyses described in this work the supervised pCCA was applied to two biological data sets and a third design data set. 
The methods can be easily extended to more than three data sets, but the determination of suitable sparsity parameters $\lambda_k$ becomes very costly.

We created an R-package, which includes functions, examples and visualizations for two biological and one design data set. It is available at http://msbi.ipb-halle.de/msbi/spCCA under GPL license.

\section{Acknowledgements}
 The gene expression experiments were funded by DFG-SPP1212 Plant Micro.

\bibliographystyle{natbib}
\bibliography{lit}

\newpage
\appendix

\section{Materials and Methods}
\subsection{Experiments}

Six different Arabidopsis lines ({\em gl1} as wildtype, single mutants {\em pen2}, {\em erp140} and {\em erp2D}, double mutants {\em pen2erp1} and {\em pen2erp2}) were used for the experiment. Plants were treated with water or {\em P. infestans} spores as described at http://bar.utoronto.ca/efp/cgi-bin/efpWeb.cgi. Leaves were harvested at 6 hours and 12 hours after inoculation. The leaves of 6 different plants per genotype were pooled for the isolation of total RNA and for metabolite profiling. The whole experiment was repeated another two times resulting in 72 samples (6 plant lines $\times$ 2 treatments $\times$ 2 time points $\times$ 3 repeats.

\subsubsection{Gene expression data}

Total RNA was isolated from Arabidopsis leaves according to \cite{Ahn09RNAextractionfrom}
and purified using the RNeasy Plant Mini Kit (Qiagen). Hybridization
of the samples to Affymetrix ATH1 GeneChips was performed by AROS
APPLIED BIOTECHNOLOGY (Aarhus, Denmark).

\subsubsection{LC/MS metabolite profiling and data processing}

Chromatographic separations were performed on an Acquity UPLC system (Waters) equipped with a HSS T3 column (100\,$\times$\,1.0\,mm, particle size 1.8\,$\mu{}$m; Waters) with a flow rate of 200\,$\mu{}$L/min at 40\,$^{\circ}\mathrm{C}$ column temperature using the following gradient program: 0 -- 60\,s, isocratic 95\% A (water/formic acid, 99.9/0.1 (v/v)), 5\% B (acetonitrile/formic acid, 99.9/0.1 (v/v)); 60 -- 360\,s, linear from 5 to 30\% B; 360 -- 600\,s, linear from 30 to 95\% B; 360 -- 720\,s isocratic 95\% B. The injection volume was 2.0\,$\mu{}\mathrm{L}$ (full loop injection). Eluted compounds were detected at a spectra rate of 3\,Hz from \textit{m/z} 100 -- 1000 using a MicrOTOF-Q-II (Bruker, Daltonics) equipped with an Apollo II electrospray ion source in positive ion mode with the following instrument settings: nebulizer gas, nitrogen, 1.2\,bar; dry gas, nitrogen, 8\,L/min, 190\,$^{\circ}\mathrm{C}$; capillary, -4500\,V; end plate offset, -500\,V; funnel 1 RF, 200\,Vpp; funnel 2 RF, 200\,Vpp. Mass calibration of individual raw data files was performed on lithium formate cluster ions obtained by automatic infusion of 20\,$\mu\mathrm{L}$ 10\,mM lithium hydroxide in isopropanol/water/formic acid, 49.9/49.9/0.2 (v/v/v) at a gradient time of 720\,s using a diverter valve.

XCMS settings for processing LC/MS data were prefilter=3,500; snthr=3; ppm=25, peakwidth=5,12. For alignment group.density function with parameters minfrac=0.75 and bw=5 was used.

\newpage
\section{Canonical variables for supervised pCCA for data assay of {\em Arabidopsis columbiana} with infection with {\em P. infestans}}

Below the eight significant variables are described and are all depicted in the figures below.
\begin{itemize}

\item First variable: Reaction to {\em P. infestans}.

\item Second variable: Mutant {\em pen2erp2}. Constantly high abundance of salicylic acid in {\em pen2erp2}-mutants. Because the infection with {\em P. infestans} was subtracted in the previous canonical variable by regression, a negative image of this variable was created, resulting in this pattern.

\item Third variable: Circadian rhythm. 
\item Forth variable: Combination of {\em pen2}-mutation and {\em erp2}-mutation. One associated gene is the {\em PEN2}-gene. {\em pen2erp2} was already subtracted in the second variable.

\item Fifth variable: Replicates and influence of different oomycete cultures on plants.
\item Sixth variable: Replicates.  

\item Seventh variable: Similarities between mutants {\em pen2erp1}, {\em pen2} and {\em erp2D}.

\item Eighth variable: Mutant {\em pen2erp1}.

\item Ninth variable: Influence of different oomycete cultures on plants.

\end{itemize}

\begin{figure}[ht!]
   
    {\includegraphics[width=0.4\textwidth]{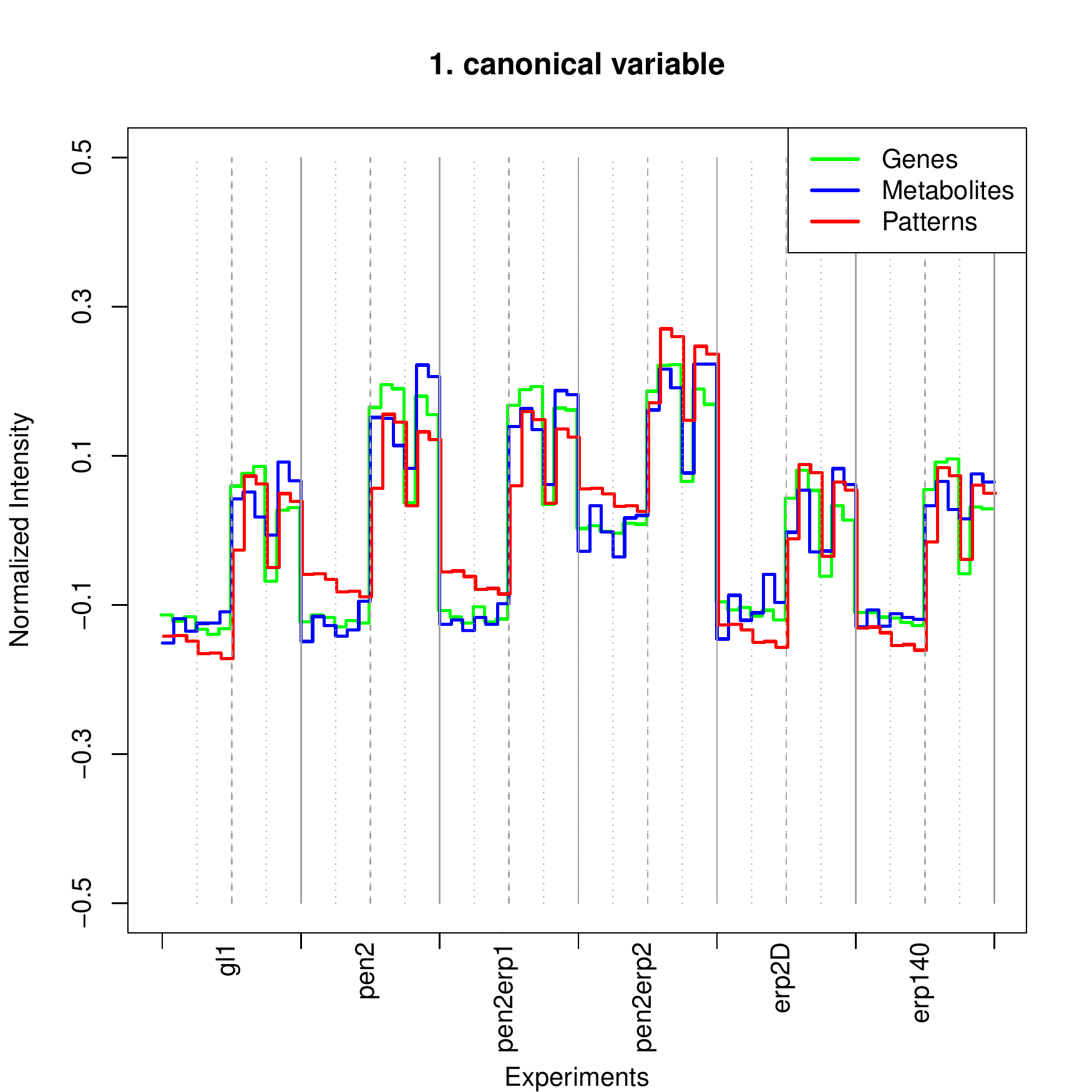}}
\end{figure}

\begin{figure}[ht!]
   
    {\includegraphics[width=0.4\textwidth]{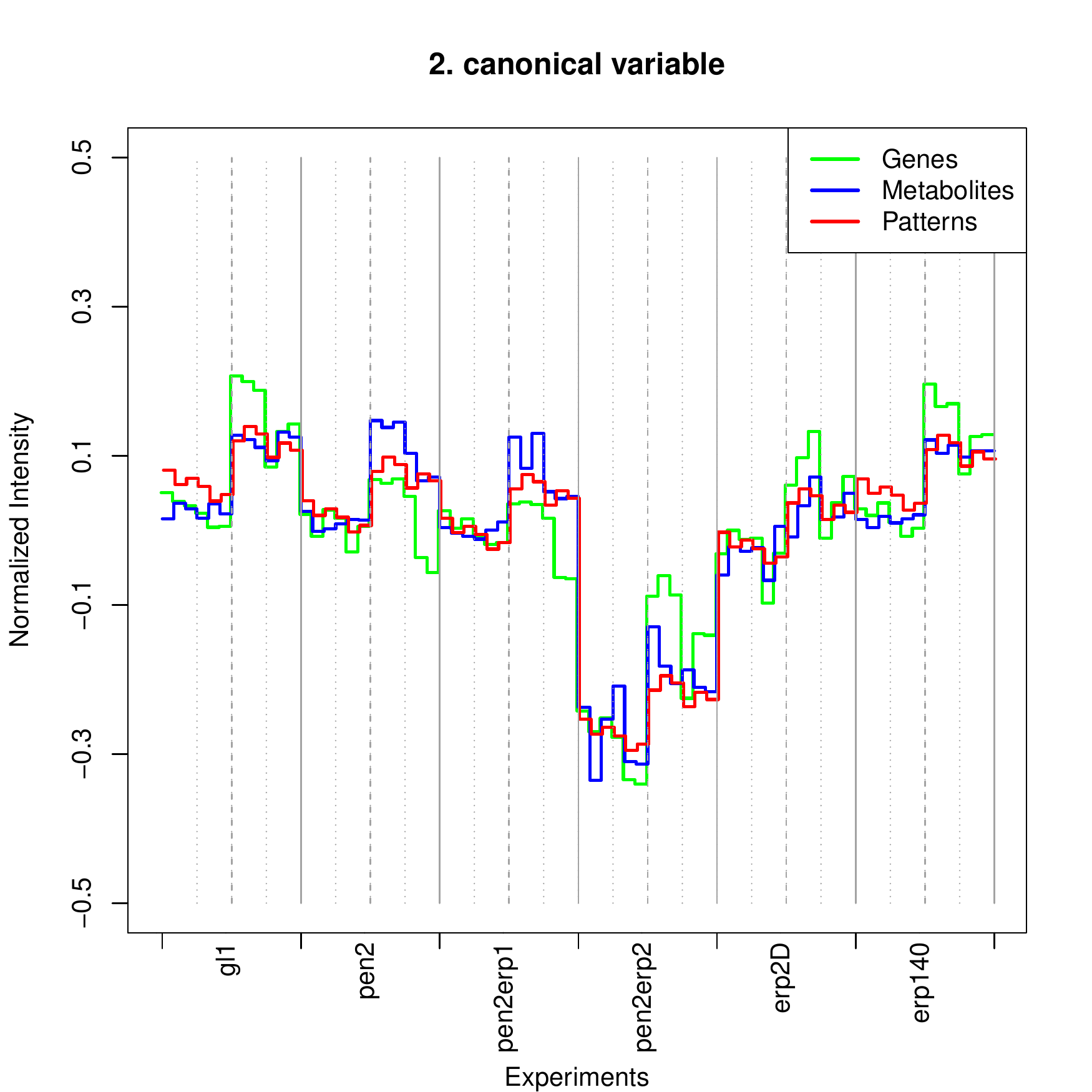}}
\end{figure}

\begin{figure}[ht!]
   
    {\includegraphics[width=0.4\textwidth]{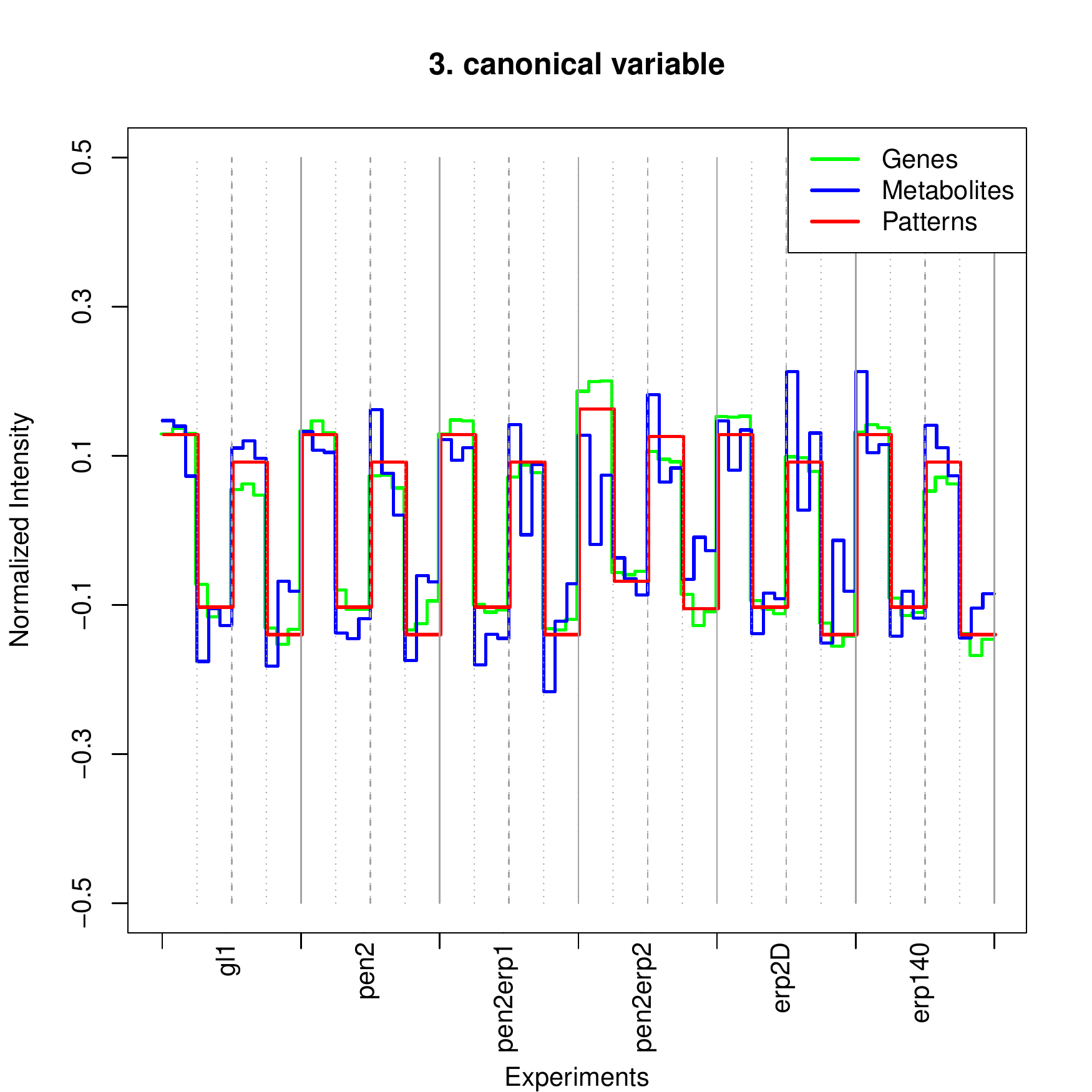}}
\end{figure}

\begin{figure}[ht!]

    {\includegraphics[width=0.4\textwidth]{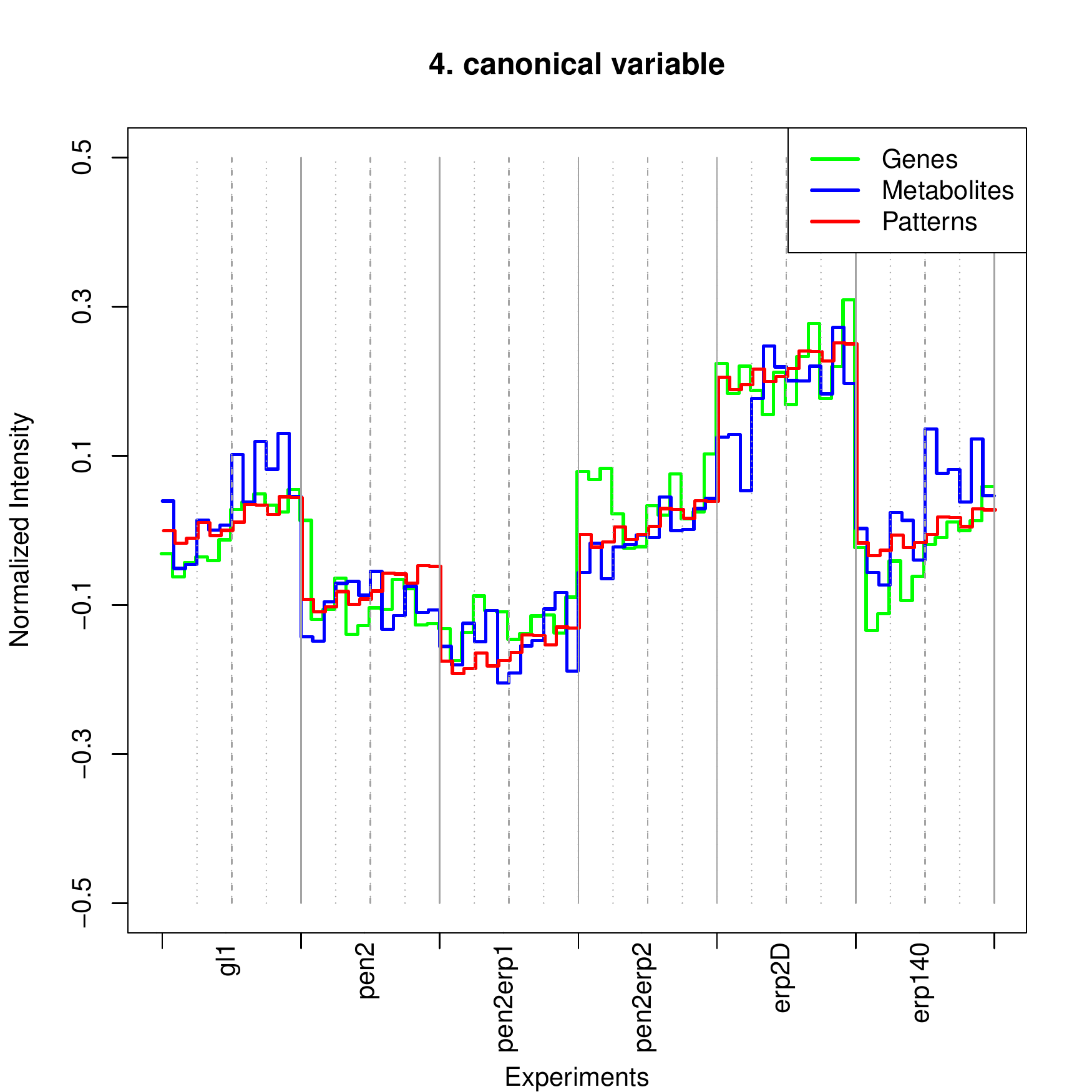}}
\end{figure}

\begin{figure}[ht!]

    {\includegraphics[width=0.4\textwidth]{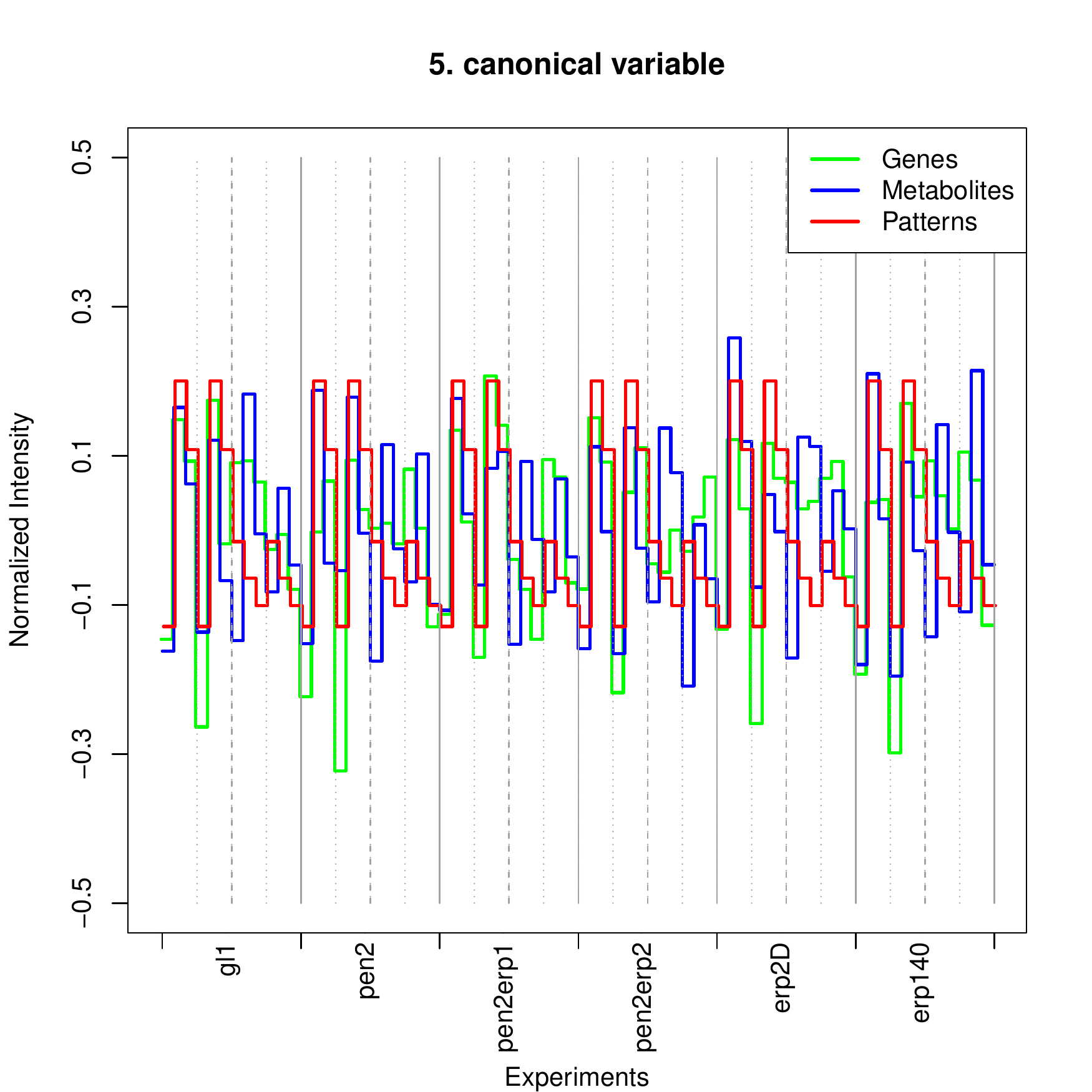}}
\end{figure}

\begin{figure}[ht!]

    {\includegraphics[width=0.4\textwidth]{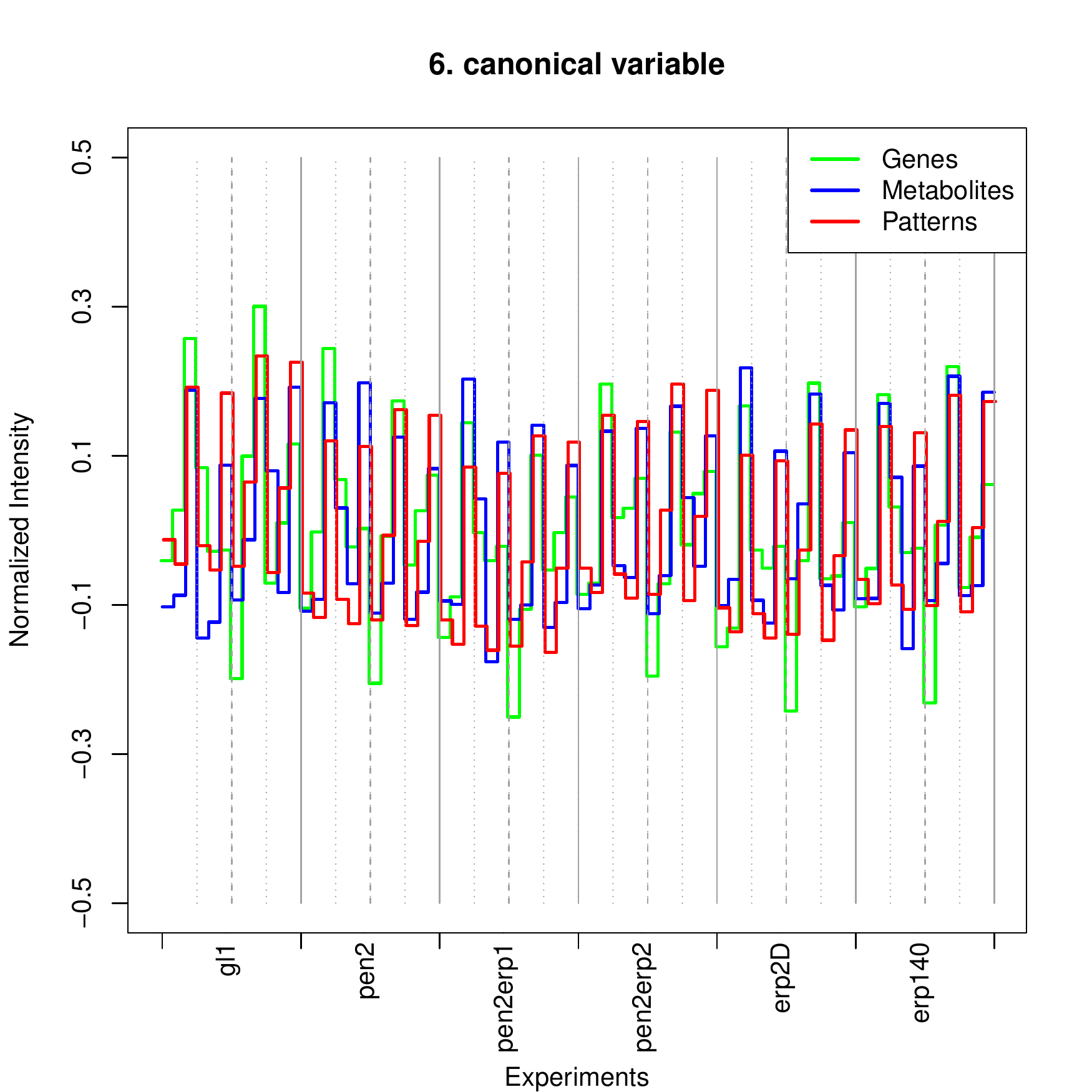}}
\end{figure}

\begin{figure}[ht!]

    {\includegraphics[width=0.4\textwidth]{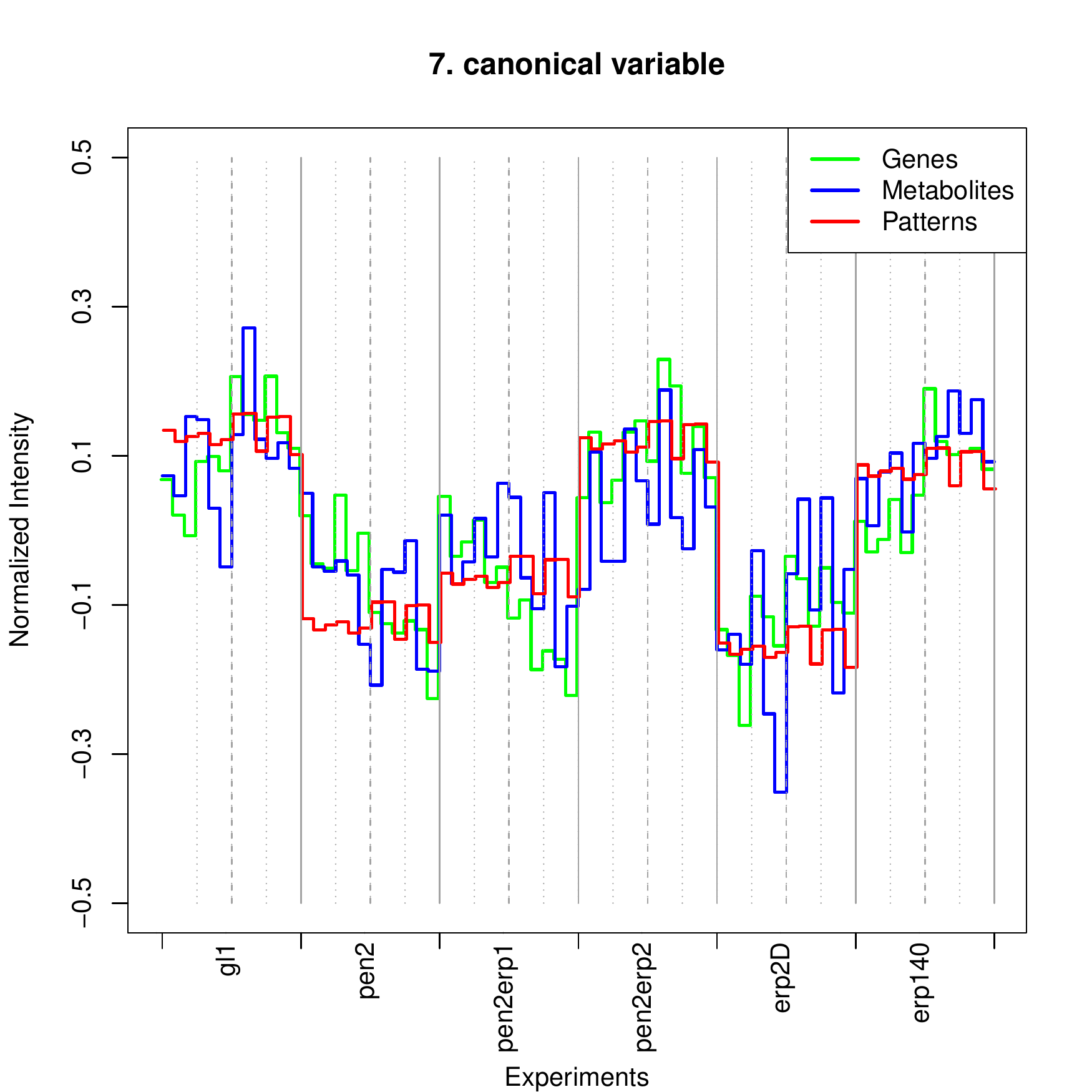}}
\end{figure}

\begin{figure}[ht!]

    {\includegraphics[width=0.4\textwidth]{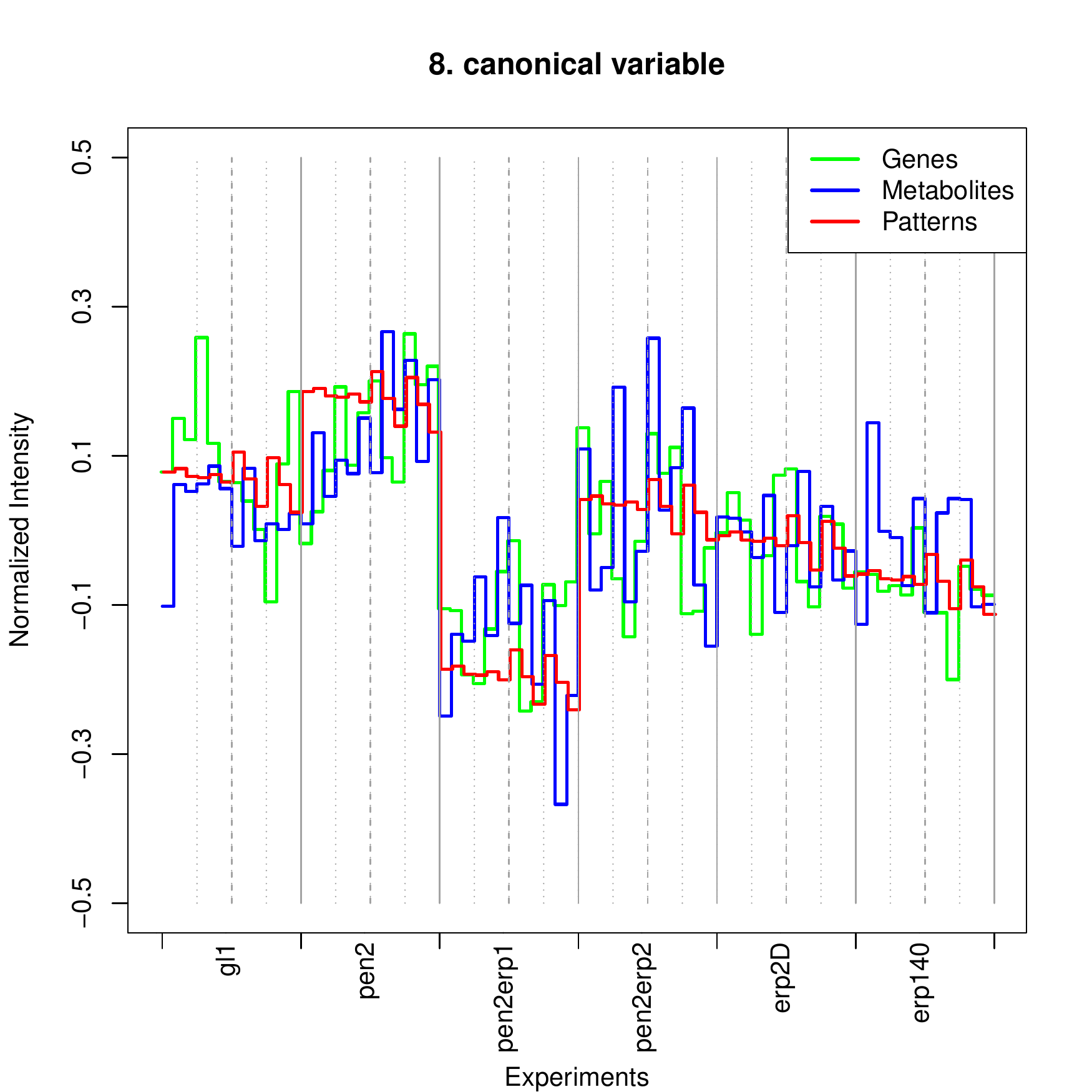}}
\end{figure}

\begin{figure}[ht!]

    {\includegraphics[width=0.4\textwidth]{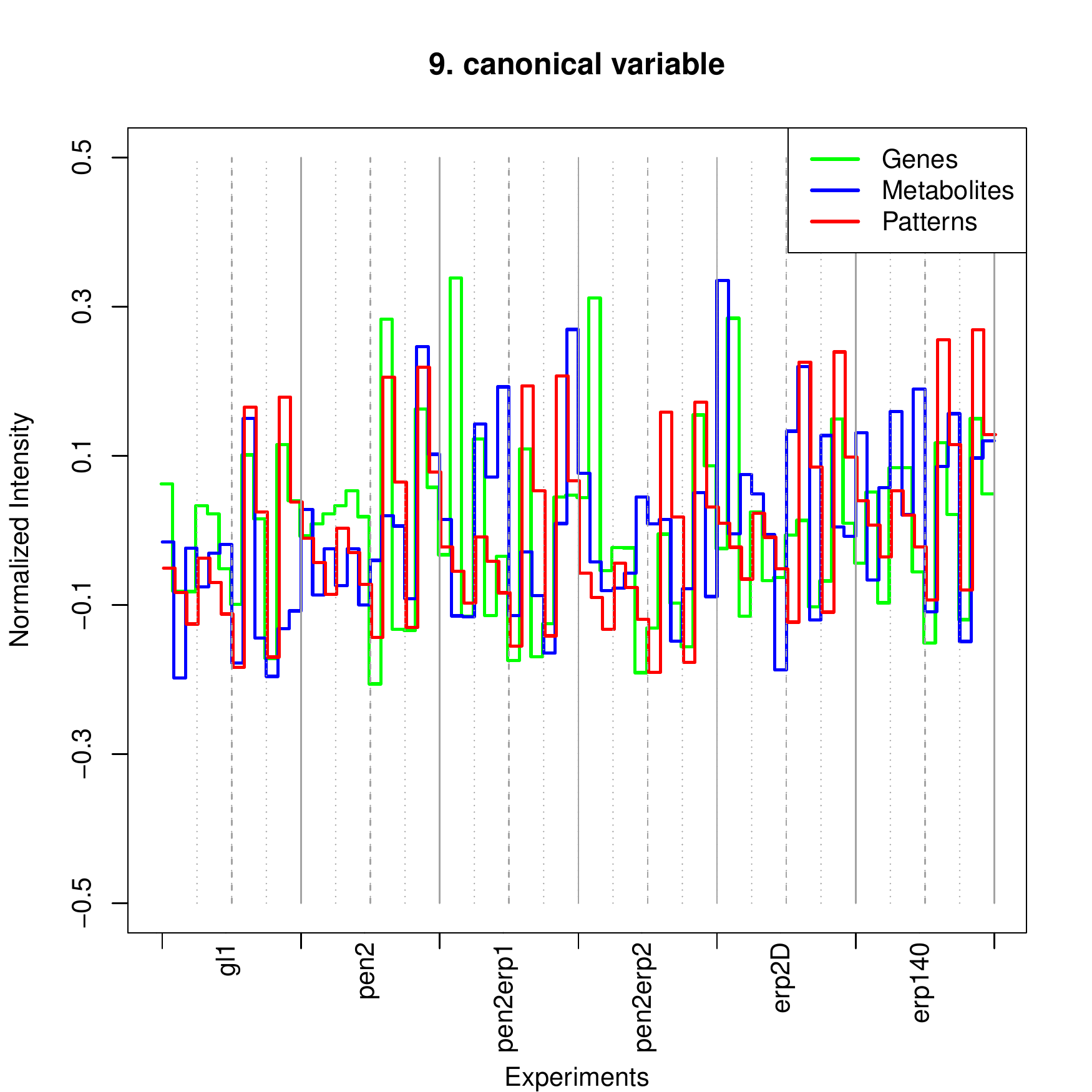}}
\end{figure}

\newpage
\section{Canonical Variables for Standard Penalized CCA}
Below, the seven significant variables are described and are all depicted in the figures below.
\begin{itemize}
\item First variable: Reaction to {\em P.infestans}, enhanced reaction for {\em pen2}-mutants. Camalexin is the main metabolite.

\item Second variable: Mutant {\em pen2erp2}. Constantly high abundance of salicylic acid in {\em pen2erp2}-mutants. Because the infection with {\em P. infestans} was subtracted in the previous canonical variable by regression, a negative image of this variable was created, resulting in this pattern.

\item Fourth variable: Circadian rhythm.

\item Further variables: Unknown processes 

\end{itemize}

\begin{figure}[ht!]
    {\includegraphics[width=0.4\textwidth]{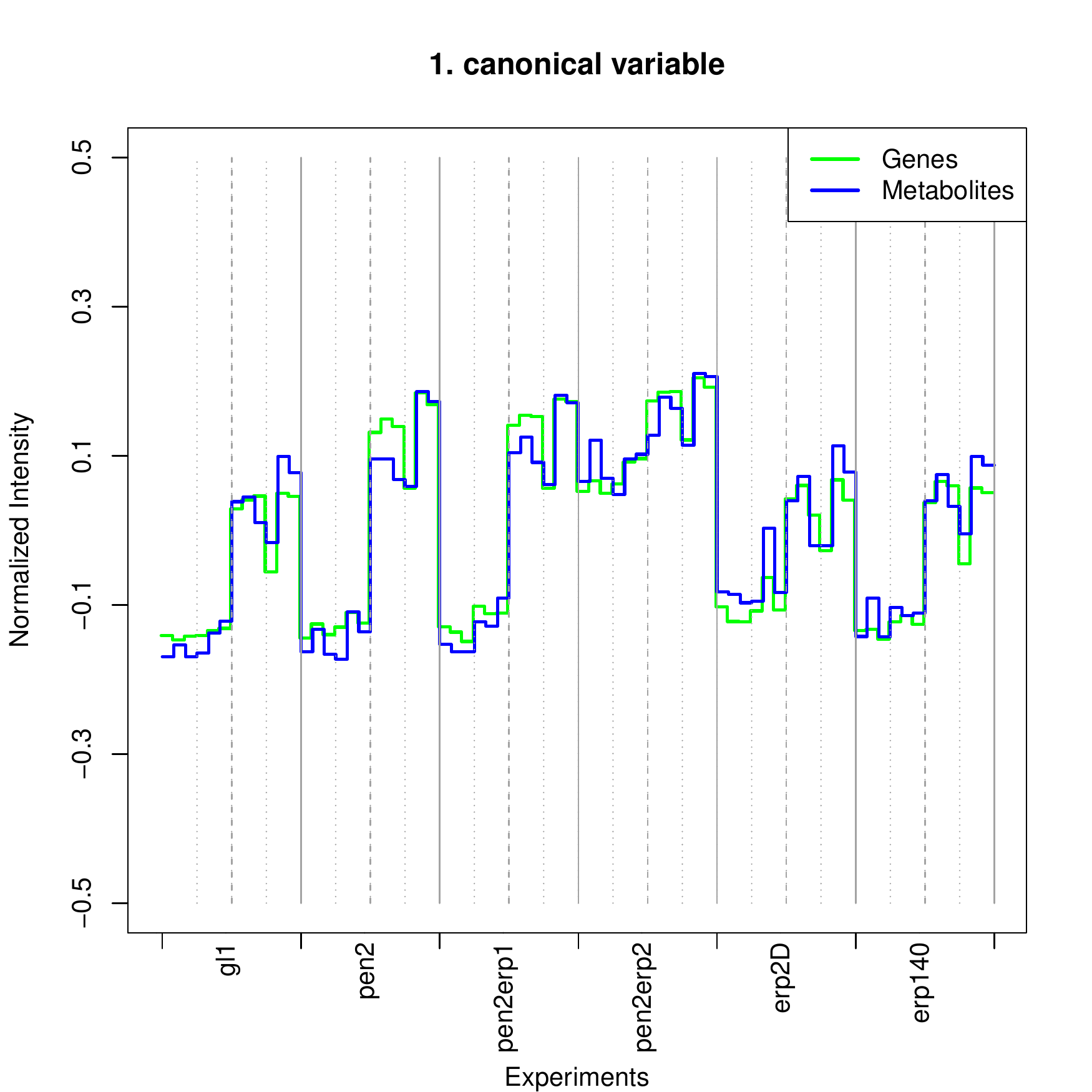}}
\end{figure}

\begin{figure}[ht!]
    {\includegraphics[width=0.4\textwidth]{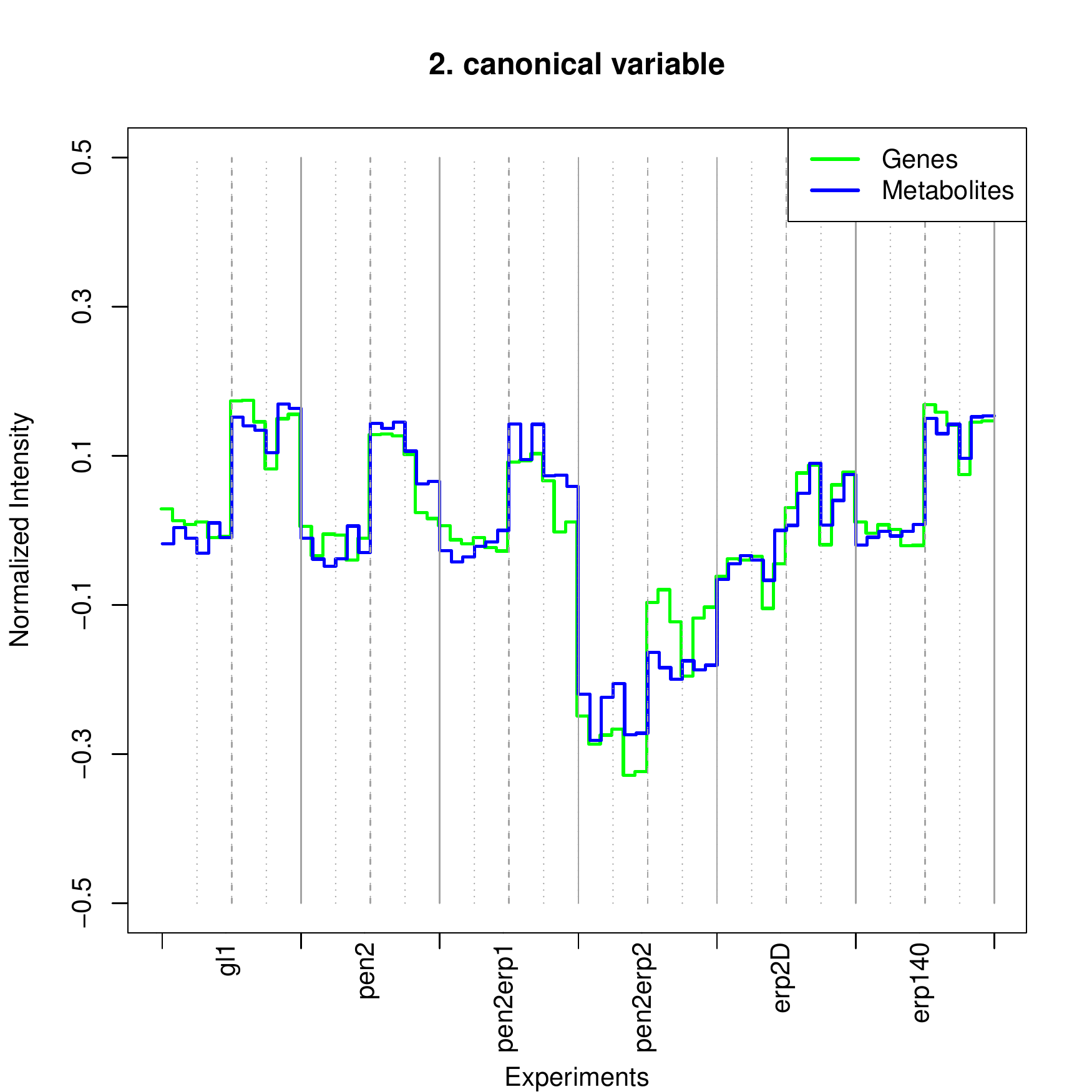}}
\end{figure}

\begin{figure}[ht!]
    {\includegraphics[width=0.4\textwidth]{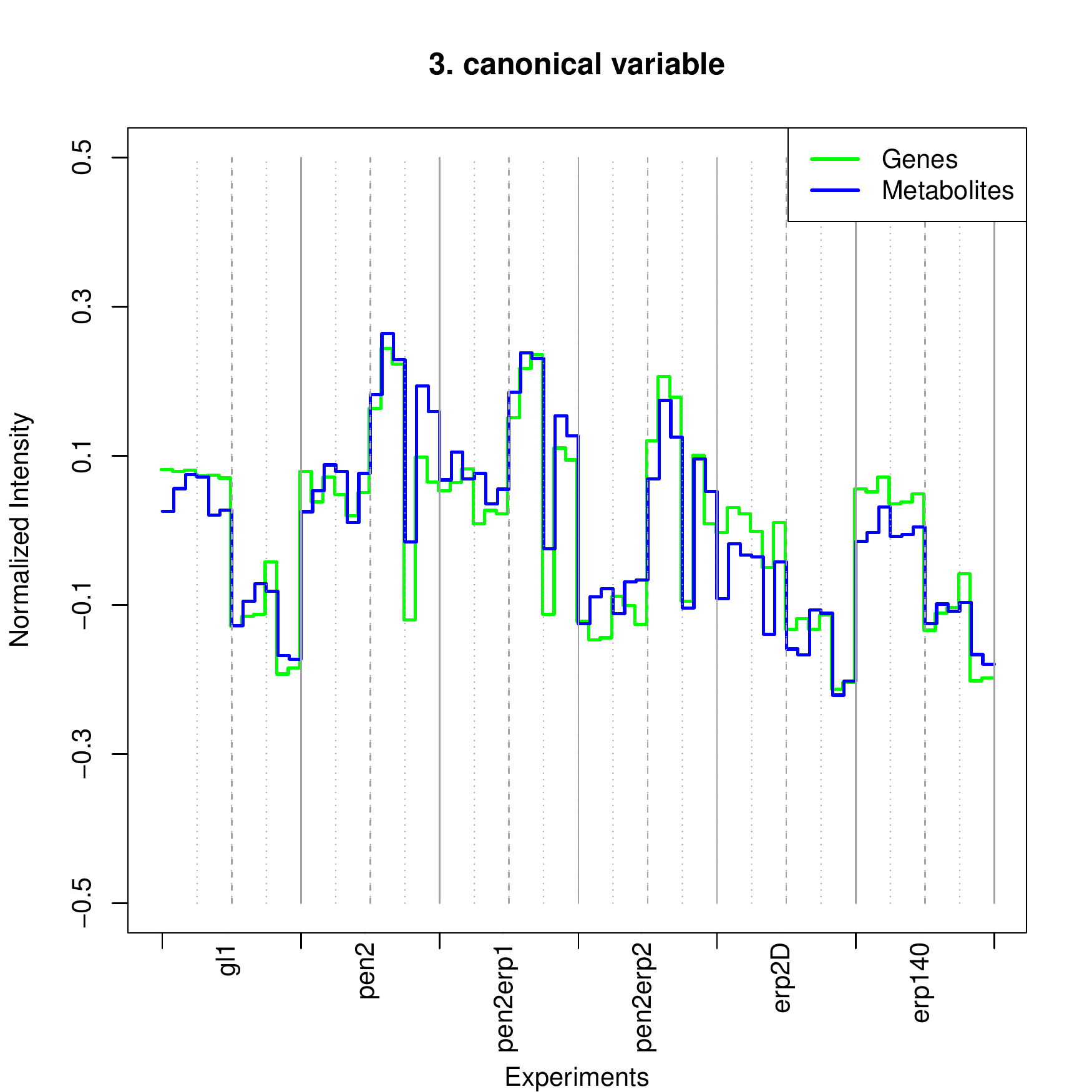}}
\end{figure}

\begin{figure}[ht!]
    {\includegraphics[width=0.4\textwidth]{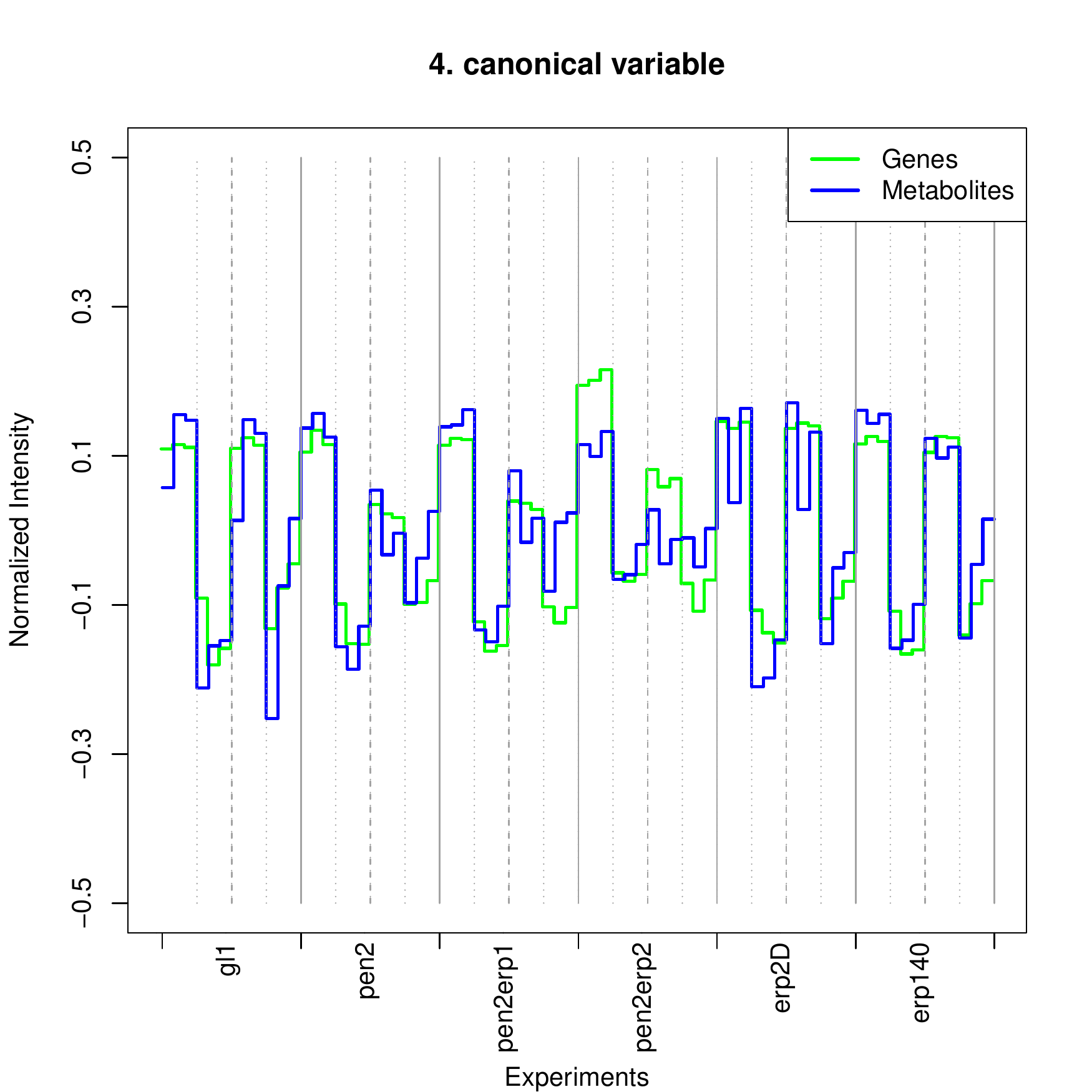}}
\end{figure}

\begin{figure}[ht!]
    {\includegraphics[width=0.4\textwidth]{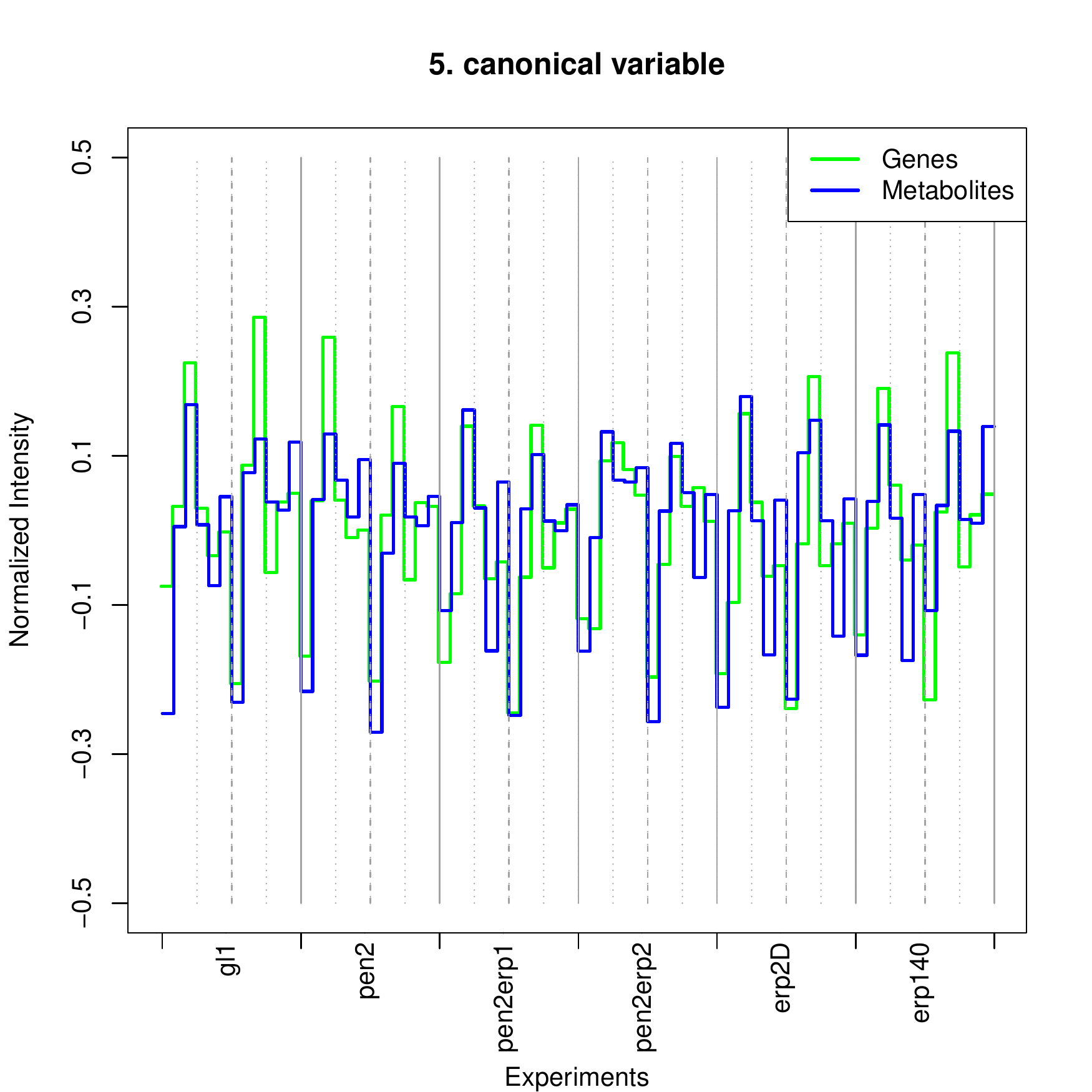}}
\end{figure}

\begin{figure}[ht!]
    {\includegraphics[width=0.4\textwidth]{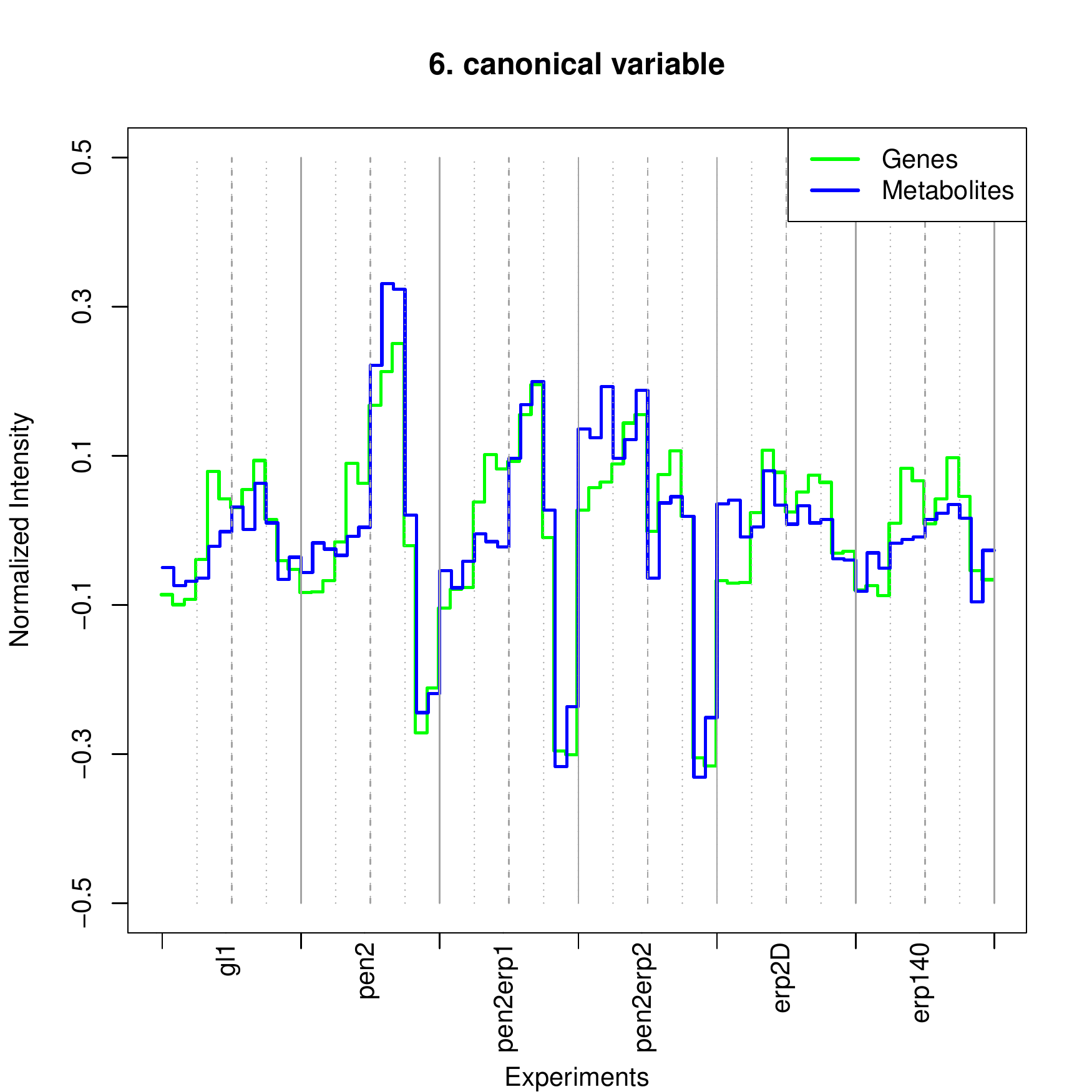}}
\end{figure}

\begin{figure}[ht!]
    {\includegraphics[width=0.4\textwidth]{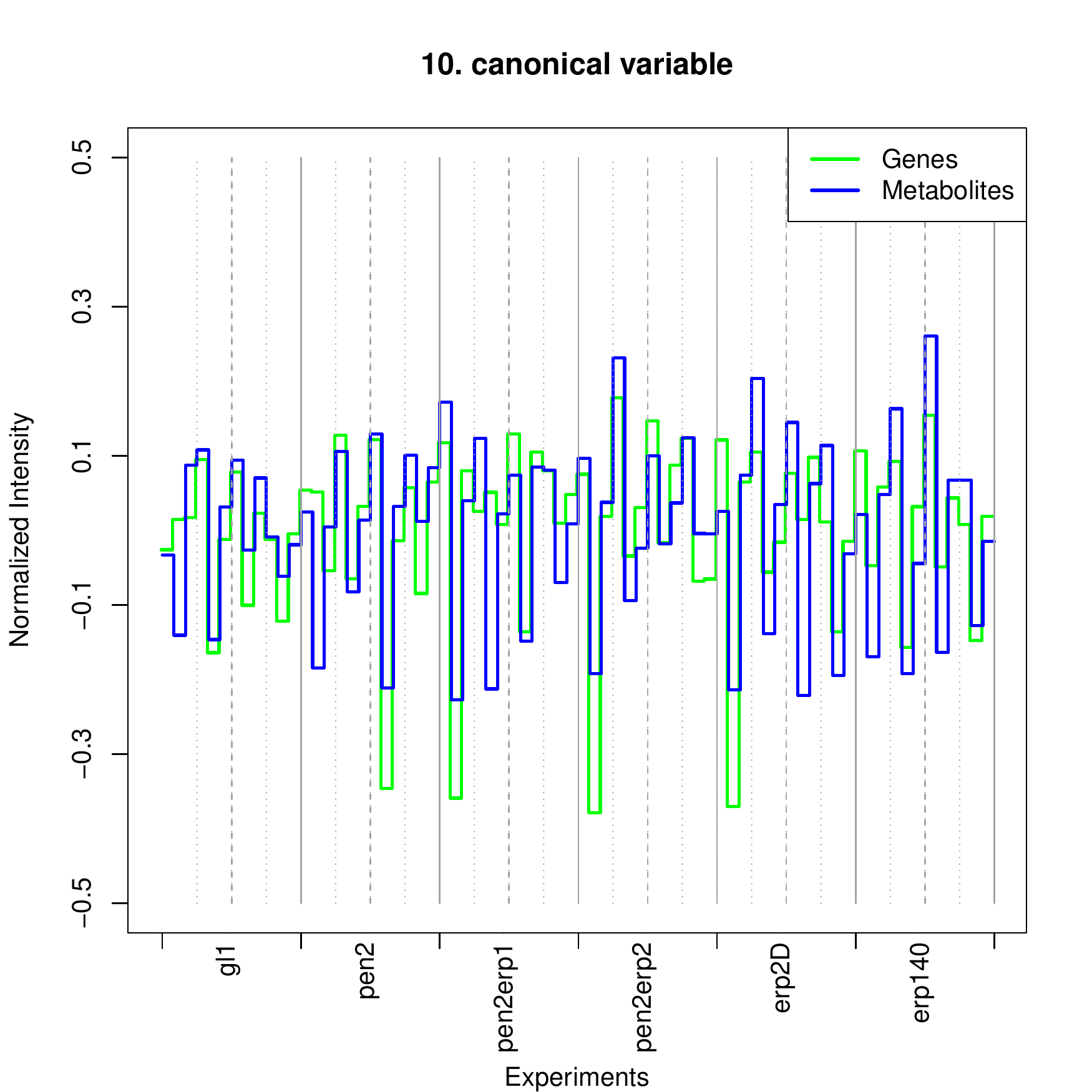}}
\end{figure}

\end{document}